\documentclass[preprint,amsmath,showpacs]{revtex4}
\def\DESepsf(#1 width #2){\epsfxsize=#2 \epsfbox{#1}}
\usepackage{graphicx}
\usepackage{color}
\usepackage{amsmath}
\def\bmatrix{\left[\begin{array}}
\def\ematrix{\end{array}\right]}

\begin{document}

%

\let\a=\alpha      \let\b=\beta       \let\c=\chi        \let\d=\delta
\let\e=\varepsilon \let\f=\varphi     \let\g=\gamma      \let\h=\eta
\let\k=\kappa      \let\l=\lambda     \let\m=\mu
\let\o=\omega      \let\r=\varrho     \let\s=\sigma
\let\t=\tau        \let\th=\vartheta  \let\y=\upsilon    \let\x=\xi
\let\z=\zeta       \let\io=\iota      \let\vp=\varpi     \let\ro=\rho
\let\ph=\phi       \let\ep=\epsilon   \let\te=\theta
\let\n=\nu
\let\D=\Delta   \let\F=\Phi    \let\G=\Gamma  \let\L=\Lambda
\let\O=\Omega   \let\P=\Pi     \let\Ps=\Psi   \let\Si=\Sigma
\let\Th=\Theta  \let\X=\Xi     \let\Y=\Upsilon

%

%

\def\cA{{\cal A}}                \def\cB{{\cal B}}
\def\cC{{\cal C}}                \def\cD{{\cal D}}
\def\cE{{\cal E}}                \def\cF{{\cal F}}
\def\cG{{\cal G}}                \def\cH{{\cal H}}
\def\cI{{\cal I}}                \def\cJ{{\cal J}}
\def\cK{{\cal K}}                \def\cL{{\cal L}}
\def\cM{{\cal M}}                \def\cN{{\cal N}}
\def\cO{{\cal O}}                \def\cP{{\cal P}}
\def\cQ{{\cal Q}}                \def\cR{{\cal R}}
\def\cS{{\cal S}}                \def\cT{{\cal T}}
\def\cU{{\cal U}}                \def\cV{{\cal V}}
\def\cW{{\cal W}}                \def\cX{{\cal X}}
\def\cY{{\cal Y}}                \def\cZ{{\cal Z}}
%

\newcommand{\Ns}{N\hspace{-4.7mm}\not\hspace{2.7mm}}
\newcommand{\qs}{q\hspace{-3.7mm}\not\hspace{3.4mm}}
\newcommand{\ps}{p\hspace{-3.3mm}\not\hspace{1.2mm}}
\newcommand{\ks}{k\hspace{-3.3mm}\not\hspace{1.2mm}}
\newcommand{\des}{\partial\hspace{-4.mm}\not\hspace{2.5mm}}
\newcommand{\desco}{D\hspace{-4mm}\not\hspace{2mm}}
\renewcommand{\figurename}{Fig.}


%
\title{\boldmath
Probing annihilations and decays of low-mass galactic dark matter in
IceCube DeepCore array: Track events}
\vfill
\author{Fei-Fan Lee}
\author{Guey-Lin Lin}
\affiliation{
 Institute of Physics,
 National Chiao-Tung University,
 Hsinchu 30010, Taiwan
}

\date{\today}
%
%
%
\begin{abstract}
The deployment of DeepCore array significantly lowers IceCube's
energy threshold to about $10$ GeV and enhances the sensitivity of
detecting neutrinos from annihilations and decays of light dark
matter. To match this experimental development, we calculate the
track event rate in DeepCore array due to neutrino flux produced by
annihilations and decays of galactic dark matter. We also calculate
the background event rate due to atmospheric neutrino flux for
evaluating the sensitivity of DeepCore array to galactic dark matter
signatures. Unlike previous approaches, which set the energy
threshold for track events at around $50$ GeV (this choice avoids
the necessity of including oscillation effect in the estimation of
atmospheric background event rate), we have set the energy threshold
at $10$ GeV to take the full advantage of DeepCore array. We compare
our calculated sensitivity with those obtained by setting 
the threshold energy at $50$ GeV. We conclude that our proposed threshold energy
significantly improves the sensitivity of DeepCore array to the dark
matter signature for $m_{\chi}< 100$ GeV in the annihilation
scenario and $m_{\chi}<300$ GeV in the decay scenario.

\end{abstract}
%
\maketitle

\pagestyle{plain}

\section{Introduction}

Many astrophysical observations have confirmed the existence of dark
matter (DM), which contributes to roughly $23\%$ of the energy
density of the Universe. Among many proposed DM candidates, weakly
interacting massive particles (WIMPs) \cite{Jungman:1996,Bertone:2005} are
popular proposals since they are theoretically well motivated and
also capable of producing the correct relic density. WIMPs could
annihilate or decay into particles such as electrons, positrons,
protons, antiprotons, photons, and neutrinos. It is possible to
establish the WIMP signature through detecting these particles
\cite{FGSTetc,INTEGRAL,WMAP-HAZE,AMS,HEAT,PAMELA,ATIC,PPB-BETS,HESS,Fermi,AMANDA,IceCube,KM3NeT}.

Research activities on WIMPs have been boosted recently in efforts
of explaining the observed anomalous positron excess in the data of
PAMELA \cite{PAMELA} and positron plus electron excess in the data
of FERMI \cite{Fermi}. To account for spectral shapes observed by
these experiments, WIMPs must annihilate or decay mostly into
leptons in order to avoid the overproduction of antiprotons. This
could indicate that DM particles are leptophilic in their
annihilations or decays \cite{leptophilic,Barger:2009}. It has been
pointed out that the observation of neutrinos can give stringent
constraints on the above scenario. Measurements of upward going
muons by Super-Kamiokande observatory place a limit on the galactic
muon neutrino flux, which in turn rules out the possibility of WIMP
annihilations to $\tau^+\tau^-$ as a source of $e^{\pm}$ anomalies
\cite{Meade:2009iu,PalomaresRuiz:2007ry,Hisano:2009}. Furthermore,
one expects that the possibilities of WIMP annihilations into
$\mu^{\pm}$, and WIMP decays into $\mu^{\pm}$ and $\tau^{\pm}$ will
all be stringently constrained
\cite{Spolyar:2009kx,Buckley:2009kw,Mandal:2009yk}(see also
discussions in Ref.~\cite{Covi:2009xn}) by the data from IceCube
detector augmented with DeepCore array.

The DeepCore array \cite{Resconi:2009,Wiebusch:2009jf} is located in
the deep center region of IceCube detector. This array consists of
$6$ densely instrumented strings plus 7 nearest standard IceCube
strings. The installation of DeepCore array significantly improves
the rejection of downward going atmospheric muons in IceCube and
lowers the threshold energy for detecting muon track or cascade
events to about $5$ GeV.  As summarized in
Ref.~\cite{Wiebusch:2009jf}, the low detection threshold of DeepCore
array is achieved by three improvements over the IceCube detector.
First, the photo-sensors in DeepCore are more densely instrumented
than those of IceCube, as just mentioned. Second, the ice surrounding the DeepCore
array is on average twice as clear as the average ice above 2000 m
\cite{optical}. Such a property is useful for reconstructing lower-
energy neutrino events. Finally the DeepCore array uses new type of
phototube which has a higher quantum efficiency.

It is clear that DeepCore array improves the sensitivity as well as
enlarges the energy window for observing neutrinos from DM
annihilations or decays in the galactic halo. Previous analyses on
the detection of these neutrinos in DeepCore
\cite{Mandal:2009yk,Erkoca:2010} have set the threshold energy at
$(40-50)$ GeV for both track and cascade events. For neutrino events
with energies higher than $50$ GeV, the estimation of atmospheric
background event rate is straightforward since oscillation effects
can be neglected. However, to take the full advantage of DeepCore
array, it is desirable to estimate the track and shower event rates
due to atmospheric neutrinos in the energy range $10 \ {\rm GeV}\leq
E_{\nu}\leq 50 \ {\rm GeV}$. In this energy range, the oscillations
of atmospheric neutrinos cannot be neglected. In this article, we
take into account this oscillation effect and calculate the track
event rate with a threshold energy $E_{\mu}^{\rm th}=10$ GeV due to
atmospheric muon neutrinos from all zenith angles. Given such a
background event rate, we then evaluate the sensitivities of DeepCore
array to the neutrino flux arising from DM annihilations and decays
in the galactic halo. In the subsequent paper, we shall analyze the
corresponding sensitivities associated with cascade events.

This paper will focus on neutrino signature induced by low-mass DM.
Hence our interested DM mass range is far below TeV level implied by
PAMELA and FERMI data. Therefore we shall consider neutrino flux
induced by DM annihilations/decays into both leptons and hadrons.
Specifically, we consider the channels $\chi\chi\to b\bar{b}, \
\tau^+ \tau^-$, and $\mu^+\mu^-$ for annihilations and the channels
$\chi\to b\bar{b}, \ \tau^+ \tau^-$ and $\mu^+\mu^-$ for decays.
Since we are only interested in low-mass dark matter, we have
neglected neutrino fluxes generated through DM annihilations or
decays into $t\bar{t}$, $W^+W^-$ and $ZZ$ final states. We also
neglect neutrino fluxes arising from light meson decays, as the
annihilation cross section for $\chi\chi\to q\bar{q}$ is likely to
be suppressed by $m_q^2$ \cite{Bertone:2005}. We shall compare the
constraints on DM annihilation cross section and DM decay time for
different values of threshold energy $E_{\mu}^{\rm th}$. For such a
comparison, we employ the modes $\chi\chi\to \mu^+\mu^-$ and
$\chi\to \mu^+\mu^-$ for illustrations.

This paper is organized as follows. In Sec. II, we outline the
calculation of muon neutrino flux from WIMP annihilations and decays
in the galactic halo. In Sec. III, we calculate the atmospheric muon
neutrino flux from all zenith angles with $E_{\nu}\geq 10$ GeV. The
oscillations between $\nu_{\mu}$ and $\nu_{\tau}$ are taken into
account. In Sec. IV, we evaluate the sensitivity of DeepCore array
to neutrino flux arising from WIMP annihilations or decays in the
galactic halo. We compare our results with those obtained by setting
$E_{\mu}^{\rm th}=50$ GeV. We summarize in Sec. V.

\section{Neutrino Flux from Annihilations and Decays of Dark Matter in the Galactic Halo}

The differential neutrino flux from the galactic dark matter halo
for neutrino flavor $i$ can be written as~\cite{Hisano:2009}
\begin{equation}
\frac{\mbox{d}\Phi_{\nu_{i}}}{\mbox{d}E_{\nu_{i}}}=
\frac{\Delta\Omega}{4\pi}\frac{\langle\sigma\upsilon\rangle}{2m^{2}_{\chi}}
\left(\sum_F B_{F}\frac{\mbox{d}N^{F}_{\nu_{i}}}{\mbox{d}E}\right)
R_{\odot}\rho^{2}_{\odot}\times J_{2}(\Delta\Omega) \\
\label{eq1}
\end{equation}
for the case of annihilating DM, and
\begin{equation}
\frac{\mbox
{d}\Phi_{\nu_{i}}}{\mbox{d}E_{\nu_{i}}}=\frac{\Delta\Omega}{4\pi}\frac{1}{m_{\chi}\tau_{\chi}}
\left(\sum_F B_{F}\frac{\mbox{d}N^{F}_{\nu_{i}}}{\mbox{d}E}\right)
R_{\odot}\rho_{\odot}\times J_{1}(\Delta\Omega) \\
\label{eq2}
\end{equation}
for the case of decaying DM, where $R_{\odot} = 8.5~\textrm{kpc}$ is
the distance from the galactic center (GC) to the solar system,
$\rho_{\odot}=0.3~\textrm{GeV}/\textrm{cm}^{3}$ is the DM density in
the solar neighborhood, $m_{\chi}$ is the DM mass, $\tau_{\chi}$ is
the DM decay time and $\mbox{d}N^{F}_{\nu_{i}}/\mbox{d}E$ is the
neutrino spectrum per annihilation or decay for a given annihilation
or decay channel $F$ with a corresponding branching fraction
$B_{F}$. The neutrino spectra $\mbox{d}N^{F}_{\nu_{i}}/\mbox{d}E$
for different channels are summarized in Ref~\cite{Erkoca:2010,Reno:2010}. The
quantity $\langle\sigma\upsilon\rangle$ is the thermally averaged
annihilation cross section, which can be written as
\begin{equation}
\langle\sigma\upsilon\rangle = B \langle\sigma\upsilon\rangle_{0},           \\
\label{eq3}
\end{equation}
with a boost factor $B$ \cite{Sommerfeld,Boost}. We set
$\langle\sigma\upsilon\rangle_{0}=3\times10^{-26}~\textrm{cm}^{3}\textrm{s}^{-1}$,
which is the typical annihilation cross section for the present dark
matter abundance under the standard thermal relic scenario
\cite{Jungman:1996}. We treat the boost factor $B$ as a
phenomenological parameter. The dimensionless quantity
$J_{n}(\Delta\Omega)$ is the DM distribution integrated over the
line-of-sight (l.o.s) and averaged over a solid angle
$\Delta\Omega=2\pi(1 - \textrm{cos}\psi_{\textrm{max}} )$, i.e.,
\begin{equation}
J_{n}(\Delta\Omega) =
\frac{1}{\Delta\Omega}\int_{\Delta\Omega}\mbox{d}\Omega\int_{\rm{l.o.s}}\frac{\mbox{d}l}{R_{\odot}}
\left(\frac{\rho(r(l,\psi))}{\rho_{\odot}}\right)^{n},           \\
\label{eq4}
\end{equation}
\\
where $\rho$ is the DM density at a specific location described by
the coordinate $(l,\psi)$ with $l$ the distance from the Earth to DM
and $\psi$ the direction of DM viewed from the Earth with $\psi=0$
corresponding to the direction of GC. The distance $r\equiv
\sqrt{R^{2}_{\odot} + l^{2} - 2R_{\odot}l\textrm{cos}\psi }$ is the
distance from GC to DM. The upper limit of the integration,
$l_{\textrm{max}} \equiv R_{\odot}\textrm{cos}\psi + \sqrt{R^{2}_{s}
- R^{2}_{\odot} \textrm{sin}^{2} \psi}$, depends on $R_s$, the
adopted size of the galactic halo. In this analysis, we take $R_{s}
= 20~\textrm{kpc}$ and use the Navarro-Frenk-White (NFW) DM density
profile~\cite{NFW}
\begin{equation}
\rho(r) =
\rho_{s}\left(\frac{R_{s}}{r}\right)\left(\frac{R_{s}}{R_{s} +
r}\right)^{2},
\end{equation}
\\
with $\rho_{s}=0.26~\textrm{GeVcm}^{-3}$ such that $\rho_{\odot}=0.3~\textrm{GeVcm}^{-3}$.

Neutrinos are significantly mixed through oscillations when they
travel a vast distance across the galaxy. We determine neutrino
flavor oscillation probabilities in the tribimaximal
limit~\cite{tribimax} of neutrino mixing angles, i.e.,
$\sin^2{\theta_{23}} = 1/2$, $\sin^2{\theta_{12}} = 1/3$ and
$\sin^2{\theta_{13}} = 0$. The neutrino fluxes on Earth are related
to those at the source through \cite{Learned:1994wg,AB,Lai}
\begin{equation}
\Phi_{\nu_{e}} = \frac{5}{9}\Phi_{\nu_{e}}^{0} + \frac{2}{9}\Phi_{\nu_{\mu}}^{0} + \frac{2}{9}\Phi_{\nu_{\tau}}^{0},
\end{equation}
and
\begin{equation}
\Phi_{\nu_{\mu}} = \Phi_{\nu_{\tau}} = \frac{2}{9}\Phi_{\nu_{e}}^{0}
+ \frac{7}{18}\Phi_{\nu_{\mu}}^{0} +
\frac{7}{18}\Phi_{\nu_{\tau}}^{0},
\end{equation}
where $\Phi_{\nu_{i}}^{0}$ is the neutrino flux of flavor $i$ at the
astrophysical source. It is understood that the recent T2K
~\cite{Abe:2011sj} and Double Chooz~\cite{DoubleChooz} experiments have indicated a non-zero value for
$\theta_{13}$. Taking the T2K best-fit value
$\sin^22\theta_{13}=0.11$ at the CP phase $\delta=0$ for the normal
mass hierarchy, we have
\begin{eqnarray}
\Phi_{\nu_{e}}& =&
0.53\Phi_{\nu_{e}}^{0}+0.26\Phi_{\nu_{\mu}}^{0}+0.21
\Phi_{\nu_{\tau}}^{0},\nonumber \\
\Phi_{\nu_{\mu}}& =&
0.26\Phi_{\nu_{e}}^{0}+0.37\Phi_{\nu_{\mu}}^{0}+0.37
\Phi_{\nu_{\tau}}^{0},\nonumber \\
\Phi_{\nu_{\tau}}& =&
0.21\Phi_{\nu_{e}}^{0}+0.37\Phi_{\nu_{\mu}}^{0}+0.42
\Phi_{\nu_{\tau}}^{0}.
\end{eqnarray}
To proceed our discussions, let us first take the neutrinos at the
source to be those generated by $B$ meson decays following the
$\chi\chi\to b\bar{b}$ annihilation. In this special case,
$\Phi_{\nu_{e}}^{0}=\Phi_{\nu_{\mu}}^{0}=\Phi_{\nu_{\tau}}^{0}$ at
the source, and consequently the relation
$\Phi_{\nu_{e}}=\Phi_{\nu_{\mu}}=\Phi_{\nu_{\tau}}$ always holds due
to the probability conservation, irrespective of the form of oscillation probability matrix.
Let us now consider
neutrinos produced at the source by muon decays following the
$\chi\chi\to \mu^+\mu^-$ annihilation. In this case, one has
$\Phi_{\nu_{e}}^{0}=\Phi_{\nu_{\mu}}^{0}$ and
$\Phi_{\nu_{\tau}}^{0}=0$. Taking
$\Phi_{\nu_{e}}^{0}=\Phi_{\nu_{\mu}}^{0}\equiv \Phi^0$, one obtains
$\Phi_{\nu_{e}}=0.78\Phi^0$ and $\Phi_{\nu_{\mu}} =
\Phi_{\nu_{\tau}}=0.61\Phi^0$ for tribimaximal values of neutrino
mixing parameters. On the other hand, with the T2K best-fit $\theta_{13}$
value, one arrives at $\Phi_{\nu_{e}}=0.79\Phi^0$,
$\Phi_{\nu_{\mu}}=0.63\Phi^0$ and $\Phi_{\nu_{\tau}}=0.58\Phi^0$ for
the normal mass hierarchy. Clearly $\Phi_{\nu_{e}}$ is almost
unaffected while $\Phi_{\nu_{\mu}}/\Phi_{\nu_{\tau}}-1=9\%$. For the
inverted mass hierarchy, one obtains the same $\nu_e$ flux while
$\Phi_{\nu_{\mu}}/\Phi_{\nu_{\tau}}-1=12\%$. Hence T2K result
implies an $O(10\%)$ difference between the arrival $\nu_{\mu}$ and
$\nu_{\tau}$ fluxes for neutrinos produced by $\chi\chi\to
\mu^+\mu^-$ annihilations.
\section{ATMOSPHERIC NEUTRINO FLUXES }

Knowing atmospheric neutrino background is important for evaluating
the sensitivity of DeepCore array to neutrino flux from DM
annihilations or decays. We begin by computing the flux of intrinsic
atmospheric muon neutrinos arising from pion and kaon decays,
following the approaches in Refs.~\cite{Gaisser:2001sd,Lee:2006}.
The $\nu_{\mu}$ flux arising from $\pi$ decays can be written as
\begin{eqnarray}
\label{atm-nu}
 \frac{\mbox{d}^2N^{\pi}_{\nu_{\mu}}(E,\xi,X)}{\mbox{d}E\mbox{d}X}&=&\int_E^{\infty}
 \mbox{d}E_N\int_E^{E_N}
 \mbox{d}E_{\pi}\frac{\Theta(E_{\pi}-\frac{E}{1-\gamma_{\pi}})}{d_{\pi}E_{\pi}(1-\gamma_{\pi})}
 \nonumber \\
 & &\times \int_0^X
 \frac{\mbox{d}X'}{\lambda_N}P_{\pi}(E_{\pi},X,X')
 \frac{1}{E_{\pi}}F_{N\pi}(E_{\pi},E_N)\nonumber \\
 & & \times \exp \left(-\frac{X'}{\Lambda_N}\right)\phi_N(E_N),
\end{eqnarray}
where $E$ is the neutrino energy, $X$ is the slant depth in units of
g/cm$^2$,  $\xi$ is the zenith angle in the direction of incident
cosmic-ray nucleons, $r_{\pi}=m_{\mu}^2/m_{\pi}^2$, $d_{\pi}$ is the
pion decay length in units of g/cm$^2$, $\lambda_N$ is the nucleon
interaction length and $\Lambda_N$ is the corresponding nucleon
attenuation length. The function $P_{\pi}(E_{\pi},X,X')$ is the
probability that a charged pion produced at the slant depth $X'$
survives to the depth $X$ ($> X'$), which is given by
\cite{Lipari:1993hd}
\begin{equation}
\label{pi_survive_a}
P_{\pi}(E_{\pi},X,X')=\exp\left(-\frac{X-X'}{\Lambda_{\pi}}\right)\cdot
\exp\left(-\frac{m_{\pi}c}{E_{\pi}\tau_{\pi}}\int_{X'}^X\frac{{\mbox
d}T}{\rho(T)}\right),
\end{equation}
where $\Lambda_{\pi}=160$ g/cm$^2$ is the pion attenuation length,
$\tau_{\pi}$ is the pion lifetime at its rest frame, while $\rho(T)$
is the atmosphere mass density at the slant depth $T$. Finally,
$F_{N\pi}(E_{\pi},E_N)$ is the normalized inclusive cross section
for $N+{\rm air}\to \pi^{\pm}+Y$, which is given by
\cite{Gaisser:2001sd}
\begin{equation}
\label{npi}
 F_{N\pi}(E_{\pi},E_N)\equiv
\frac{E_{\pi}}{\sigma_N}\frac{{\mbox d}\sigma(E_{\pi},E_N)}{{\mbox
d}E_{\pi}}=c_+(1-x)^{p_+}+c_-(1-x)^{p_-},
\end{equation}
with $x=E_{\pi}/E_N$, $c_+=0.92$, $c_-=0.81$, $p_+=4.1$, and
$p_-=4.8$.

The primary cosmic-ray spectrum $\phi_N(E_N)$ in Eq.~(\ref{atm-nu})
includes contributions from cosmic ray protons and those from
heavier nuclei. We have $\phi_N(E_N)=\sum_{A}A\phi_A(E_N)$ with $A$
the atomic number of each nucleus. The spectrum of each cosmic-ray
component is parametrized by \cite{Gaisser:2002jj,Honda:2004}
\begin{eqnarray}
\label{phi n}
\phi_A(E_N) = K\times(E_N + b\,\textrm{exp}[- c
\sqrt{E_N}])^{- \alpha},
\end{eqnarray}
in units of m$^{-2}$s$^{-1}$sr$^{-1}$GeV$^{-1}$. The fitting
parameters $\alpha,K,b,c$ depend on the type of nucleus. They are
tabulated in Table I \cite{Honda:2004}. The kaon contribution to the
atmospheric $\nu_{\mu}$ flux has the same form as Eq.~(\ref{atm-nu})
with an inclusion of the branching ratio $B(K\to \mu\nu)=0.635$ and
appropriate replacements in kinematic factors and the normalized
inclusive cross section. In particular, $F_{NK}(E_{K},E_N)$ can be parametrized as  Eq.~(\ref{npi}) with $c_{+}=0.037$, $c_{-}=0.045$, $p_{+}=0.87$, and $p_-=3.5$.
\begin{table}
\caption{Parameters for all five components in the fit of Eq.~(\ref{phi n}).}
\begin{center}
\begin{ruledtabular}
\begin{tabular}{ccccc}
parameter/component & $\alpha$ & $K$ & $b$ & $c$
\\ \hline \\
Hydrogen (A=1) $(\leq10^{2}\textrm{GeV})$ & 2.74  & 14900  & 2.15 & 0.21    \\
Hydrogen (A=1) $(>10^{2}\textrm{GeV})$ & 2.71  & 14900  & 2.15 & 0.21    \\
He (A=4) & 2.64  & 600  & 1.25 & 0.14    \\
CNO (A=14) & 2.60  & 33.2  & 0.97 & 0.01    \\
Mg-Si (A=25) & 2.79  & 34.2  & 2.14 & 0.01    \\
Iron (A=56) & 2.68  & 4.45  & 3.07 & 0.41    \\
\end{tabular}
\end{ruledtabular}
\end{center}
\end{table}

Since our interested energy range is as low as $10$ GeV, the
three-body muon decay contribution to the atmospheric $\nu_{\mu}$ flux
is not negligible, particularly in the near horizontal direction. To
obtain this part of contribution, we first compute the atmospheric
muon flux from pion and kaon decays. The muon flux induced by pion decays is given by \cite{Gaisser:2001sd,Lee:2006}
\begin{eqnarray}
\label{atm-mu}
\frac{\mbox{d}N^{\pi}_{\mu}(E,\xi,X)}{\mbox{d}E}&=&\int_{E'}^{\infty}
 \mbox{d}E_N\int_{E'}^{E_N}
 \mbox{d}E_{\pi}\int_0^{X}{\mbox
 d}X^{''}P_{\mu}(E,X,X^{''})\nonumber \\
 &\times& \frac{\Theta(E_{\pi}-E')\Theta(\frac{E'}{r_{\pi}}-E_{\pi})}{d_{\pi}E_{\pi}(1-r_{\pi})}\nonumber \\
 &\times& \int_0^{X^{''}}
 \frac{\mbox{d}X'}{\lambda_N}
  P_{\pi}(E_{\pi},X^{''},X')
  \frac{1}{E_{\pi}}F_{N\pi}(E_{\pi},E_N)\nonumber \\
 &\times& \exp \left(-\frac{X'}{\Lambda_N}\right)\phi_N(E_N),
\end{eqnarray}
where $E'$ and $E$ are muon energies at slant depths $X^{''}$ and
$X$ respectively, while $P_{\mu}(E,X,X^{''})$ is the muon survival
probability. The muon flux induced by kaon decays can be calculated in a similar way. Since $\mu^-(\mu^+)$ produced by $\pi^-(\pi^+)$ decays
are polarized, we classify muon flux into four different components
such as $\mbox{d}N^{\pi^+}_{\mu^{+}_R}/\mbox{d}E$,
$\mbox{d}N^{\pi^-}_{\mu^{-}_R}/\mbox{d}E$,
$\mbox{d}N^{\pi^+}_{\mu^{+}_L}/\mbox{d}E$, and
$\mbox{d}N^{\pi^-}_{\mu^{-}_{L}}/\mbox{d}E$. We also calculate
additional four components of the muon flux arising from the kaon
decays. Hence the $\nu_{\mu}$ flux arising
from muon decays can be written as \cite{Lipari:1993hd,Lee:2006}
\begin{eqnarray}
\label{nu_mue}
\frac{\mbox{d}^2N^{\mu^{\pm}}_{\nu_{\mu}}(E,\xi,X)}{\mbox{d}E\mbox{d}X}&=&
\sum_{s=L,R}\int_E^{\infty}{\mbox d}E_{\mu}\frac{F_{\mu^{\pm}_s\to
\nu_{\mu}}(E/E_{\mu})}{d_{\mu}(E_{\mu},X)E_{\mu}}\nonumber \\
&\times& \frac{\mbox{d}N_{\mu_s^{\pm}}(E_{\mu},\xi,X)}{\mbox{d}E_{\mu}},
\end{eqnarray}
where $d_{\mu}(E_{\mu},X)$ is the muon decay length in units of
g/cm$^2$ at the slant depth $X$ and
$F_{\mu^{\pm}_s\to\nu_{\mu}}(E/E_{\mu})$ is the normalized decay
spectrum of $\mu^{\pm}_s\to\nu_{\mu}$. Summing the two-body and
three-body decay contributions, we obtain the total intrinsic
atmospheric muon neutrino flux. In Fig.~1, we show the comparison of
angle-averaged atmospheric muon neutrino flux obtained by our
calculation and that obtained by Honda et al. \cite{Honda:2007}. At
$E_{\nu}=$ 10 GeV, two calculations only differ by $3\%$. At
$E_{\nu}=$ 100 GeV, the difference is $10\%$. We also show in the
same figure the atmospheric muon neutrino flux measured by AMANDA-II
detector \cite{Amanda:2010}. It is seen that both calculations agree
well with AMANDA results.

To completely determine the atmospheric muon neutrino flux, one also
needs to calculate the intrinsic atmospheric tau neutrino flux,
although this part of contribution is rather small. The intrinsic
atmospheric $\nu_{\tau}$ flux arises from $D_s$ decays. This flux can be obtained
by solving cascade equations
\cite{Gaisser:1992vg,Lee:2006}. We obtain
\begin{equation}
\label{atm-tau}
 \frac{\mbox{d}^2N_{\nu_{\tau}}(E,X)}{\mbox{d}E\mbox{d}X}=
 \frac{Z_{ND_s}Z_{D_s\nu_{\tau}}}{1-Z_{NN}}\cdot
 \frac{\exp(-X/\Lambda_N)\phi_N(E_N)}{\Lambda_N},
 \end{equation}
where $Z_{NN}\equiv 1-\lambda_N/\Lambda_N$ and $Z_{ND_s}$ is a special case of the generic expression
\begin{equation}
 Z_{ij}(E_j)\equiv \int_{E_j}^{\infty}{\mbox
 d}E_i\frac{\phi_i(E_i)}{\phi_i(E_j)}\frac{\lambda_i(E_j)}{\lambda_i(E_i)}
 \frac{{\mbox d}n_{iA\to jY}(E_i,E_j)}{{\mbox d}E_j},
\label{z-moment}
\end{equation}
with ${\mbox d}n_{iA\to jY}(E_i,E_j)\equiv {\mbox d}\sigma_{iA\to
jY}(E_i,E_j)/\sigma_{iA}(E_i)$ and $\lambda_i$ the interaction
length of particle $i$ in units of g/cm$^2$. The decay moment
$Z_{D_s\nu_{\tau}}$ is given by
\begin{equation}
 Z_{D_s\nu_{\tau}}(E_{\nu_{\tau}})\equiv \int_{E_{\nu_{\tau}}}^{\infty}{\mbox
 d}E_{D_s}\frac{\phi_{D_s}(E_{D_s})}{\phi_{D_s}(E_{\nu_{\tau}})}\frac{d_{D_s}
 (E_{\nu_{\tau}})}{d_{D_s}(E_{D_s})}
 F_{D_s\to \nu_{\tau}}(E_{\nu_{\tau}}/E_{D_s}),
\label{decay-moment}
\end{equation}
where $d_{D_s}$ is the decay length of $D_s$ and $F_{D_s\to
\nu_{\tau}}(E_{\nu_{\tau}}/E_{D_s})$ is the normalized decay
distribution. In this work, we employ the next-to-leading order
(NLO) perturbative QCD \cite{Nason:1989zy} with CTEQ6 parton
distribution functions to calculate the differential cross section
of $NA\to c\bar{c}$ and determine $Z_{ND_s}$.

\begin{figure}
  \centering
  \includegraphics[width=0.85\textwidth]{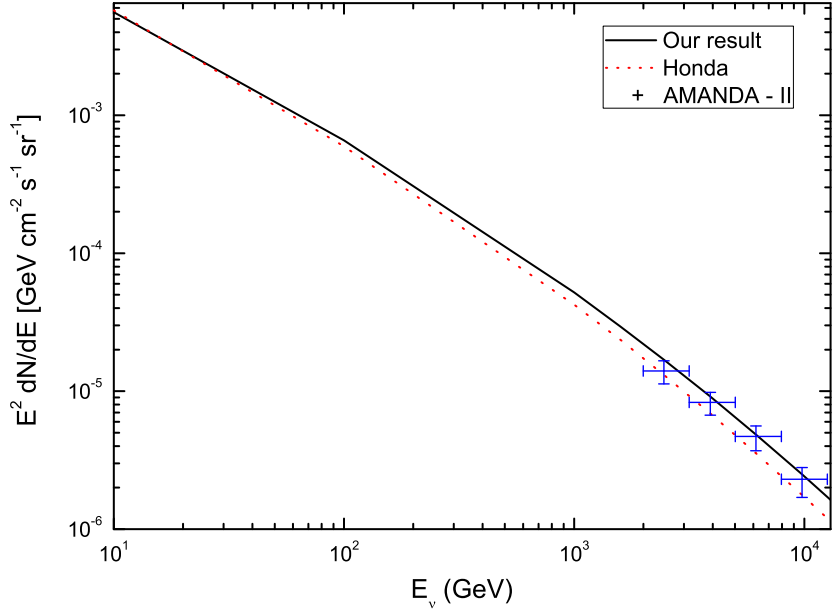}
  \vskip 0.5cm
  \caption{ The comparison of angle-averaged atmospheric muon
  neutrino ($\nu_{\mu} + \overline{\nu}_{\mu}$) flux obtained by our calculation and that obtained
  by Honda {\em et al.}~\cite{Honda:2007}. Angle-averaged $\nu_{\mu} + \overline{\nu}_{\mu}$ flux
  from AMANDA-II measurements \cite{Amanda:2010} is also shown.
  }
  \label{fig;1}
\end{figure}

Finally, the atmospheric $\nu_{\mu}$ flux taking into account the
neutrino oscillation effect is given by
\begin{eqnarray}
\label{oscillate}
\frac{\mbox{d}\bar{N}_{\nu_{\mu}}(E,\xi)}{\mbox{d}E}&=&\int
{\mbox d}X\left[
\frac{\mbox{d}^2N_{\nu_{\tau}}}{\mbox{d}E\mbox{d}X}\cdot
P_{\nu_{\tau}\to \nu_{\mu}}\right.\nonumber \\
&&\left.+\frac{\mbox{d}^2N_{\nu_{\mu}}}{\mbox{d}E\mbox{d}X}
\cdot \left(1-P_{\nu_{\mu}\to
\nu_{\tau}}\right)\right],
\end{eqnarray}
where $P_{\nu_{\mu}\to
\nu_{\tau}}\left(E,L(X,\xi)\right)=P_{\nu_{\tau}\to \nu_{\mu}}
\left(E,L(X,\xi)\right)\equiv\sin^22\theta_{23}\sin^2(1.27\Delta
m_{31}^2L/E)$ is the $\nu_{\mu}\to \nu_{\tau}$ oscillation
probability and $L(X,\xi)$ is the linear distance from the neutrino
production point to the position of IceCube DeepCore array. The unit
of $\Delta m_{31}^2$ is eV$^2$ while $L$ and $E$ are in units of km
and GeV respectively. The best-fit values for oscillation parameters
obtained from a recent analysis \cite{GonzalezGarcia:2010er} are
$\Delta m_{31}^{2}=2.47\cdot 10^{-3}\, \, \, {\rm eV}^{2}$ and
$\sin^{2}2\theta_{23} =1$ respectively.

\section{RESULTS}

In IceCube DeepCore, the event rate for contained muons is given by
\begin{eqnarray}
\label{muevent}
\Gamma_{\mu}&=&\int_{E_\mu^{\textrm{th}}}^{E_{\textrm{max}}}{\mbox
d}E_\mu
\int_{E_\mu}^{E_{\textrm{max}}}{\mbox d}E_{\nu_{\mu}} N_{A} \rho_{\textrm{ice}} V_{\textrm{tr}} \nonumber \\
&\times&\frac{d\Phi_{\nu_{\mu}}}{dE_{\nu_{\mu}}}\cdot \frac{d\sigma_{\nu N}^{\textrm{CC}}(E_{\nu_{\mu}},E_{\mu})}{dE_{\mu}}
+(\nu \rightarrow \overline{\nu}),
\end{eqnarray}
where $\rho_{\textrm{ice}} = 0.9\, \textrm{g}\,\textrm{cm}^{-3}$ is
the density of ice, $N_{A} = 6.022\times10^{23}\,\textrm{g}^{-1}$ is
Avogadro's number, $V_{\textrm{tr}}\approx0.04\,\textrm{km}^{3}$ is
the effective volume of IceCube DeepCore array for muon track events
\cite{Resconi:2009}, $d\Phi_{\nu_{\mu}}/dE_{\nu_{\mu}}$ is the muon
neutrino flux arrived at IceCube, $E_{\textrm{max}}$ is taken as
$m_{\chi}$ for annihilation and $m_{\chi}/2$ for decay, and
$E_\mu^{\textrm{th}}$ is the threshold energy for muon track events.
In this work, we use differential cross sections $d\sigma_{\nu
N}^{\textrm{CC}}(E_{\nu_{\mu}},E_{\mu})/dE_{\mu}$ given by Ref.~\cite{Gandhi:1996} with CTEQ6 parton distribution functions. We also
set $E_\mu^{\textrm{th}}=10$ GeV.

\begin{figure}
  \centering
  \includegraphics[width=0.85\textwidth]{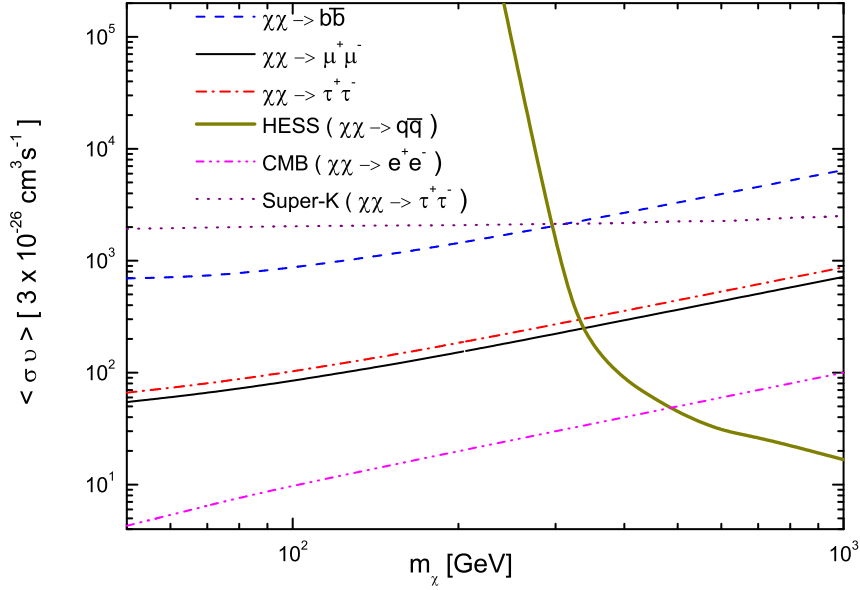}
  \vskip 0.5cm
  \caption{ 
  The dashed line, thin solid line, and dot-dashed lines are the expected constraints on DM annihilation cross section by DeepCore detector for
  $\chi\chi\rightarrow b\overline{b}$,
  $\chi\chi\rightarrow\mu^{+}\mu^{-}$, and
  $\chi\chi\rightarrow \tau^+\tau^-$ channels, respectively. The thick solid line is the H.E.S.S constraint on the annihilation cross section of DM into the light quark pair $\chi\chi\to q\bar{q}$~\cite{HESS:2011}.
  The dot-dot-dashed line is the constraint on $\chi\chi\to e^+e^-$ annihilation cross section from the analysis of cosmic microwave background (CMB) data ~\cite{Hisano:2011}.
  The dotted line is the $3\sigma$ constraint on the annihilation cross section of $\chi\chi\to \tau^+\tau^-$ from Super-Kamiokande data \cite{Meade:2009iu}.  }
  \label{fig:2}
\end{figure}

\begin{figure}
  \centering
  \includegraphics[width=0.85\textwidth]{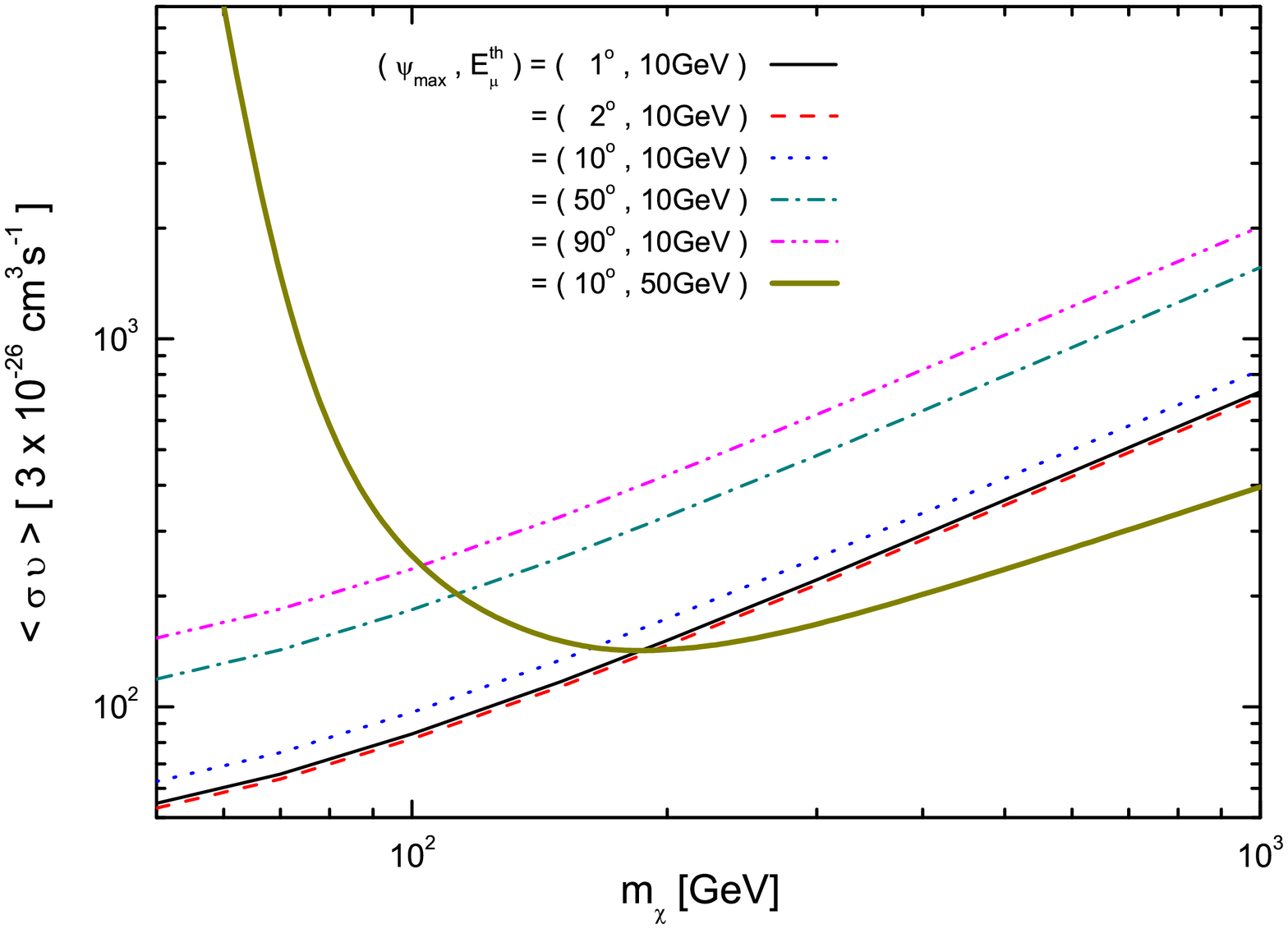}
  \vskip 0.5cm
  \caption{ The required DM annihilation cross section
  $(\chi\chi\rightarrow\mu^{+}\mu^{-})$ as a function of $m_{\chi}$
  such that the neutrino signature from DM annihilations can be detected at the $2\sigma$
  significance in five years. Results corresponding to different $\psi_{\rm max}$ are
  presented. For comparison, we also show the result with $E_{\mu}^{\rm th}=50$ GeV
  and $\psi_{\textrm{max}}=10^{\circ}$ \cite{Erkoca:2010}.
  }
  \label{fig;3}
\end{figure}
As stated before, we consider neutrino fluxes generated by the
annihilation channels $\chi\chi\to b\bar{b}, \ \tau^+ \tau^-$, and
$\mu^+\mu^-$, and the decay channels $\chi\to b\bar{b}, \ \tau^+
\tau^-$ and $\mu^+\mu^-$. Given the atmospheric neutrino background,
we present in Fig.~2 the required DM annihilation cross section as a
function of $m_{\chi}$ for threshold energy $E_\mu^{\textrm{th}}=
10~\textrm{GeV}$ and a cone half-angle
$\psi_{\textrm{max}}=1^{\circ}$ such that the neutrino signature
from DM annihilations can be detected at the $2\sigma$ significance
in five years. Non-detection of such a signature would then exclude
the parameter region above the curve at the $2\sigma$ level. We have
presented results corresponding to different annihilation channels.
One can see that the required annihilation cross section for
$2\sigma$ detection significance is smallest for
$\chi\chi\rightarrow\mu^{+}\mu^{-}$ channel and largest for the
channel $\chi\chi\rightarrow b\overline{b}$. We also present the
$3\sigma$ constraint on $\chi\chi\rightarrow\tau^{+}\tau^{-}$
annihilation cross section obtained from Super-Kamiokande data of
upward going muons~\cite{Meade:2009iu}, which has been used to rule
out WIMP annihilations into $\tau^+\tau^-$ as a possible source of
previously mentioned $e^{\pm}$ anomalies
\cite{Meade:2009iu,PalomaresRuiz:2007ry,Hisano:2009}. Such a
constraint can be compared with the expected $2\sigma$ constraint on the same
annihilation channel from DeepCore detector.

Constraints on DM annihilation cross section were also obtained from
gamma ray observations and cosmology. H.E.S.S. telescope performed a
search for the very-high-energy ($\geq 100$ GeV) $\gamma$-ray signal
from DM annihilations over a circular region of radius $1^{\circ}$
centered at the GC~\cite{HESS:2011}. With DM particles assumed to
annihilate into $q\bar{q}$ pairs, the limit on DM annihilation cross
section as a function of $m_{\chi}$ for NFW DM density profile is
derived in Ref. ~\cite{HESS:2011}. We present this constraint in
Fig.~2 as well. For $m_{\chi}>300$ GeV, the parameter space
with $B>100$ (i.e., $\langle \sigma v \rangle > 3\times 10^{-24}$m$^3$s$^{-1}$) in Fig.~2 could be excluded by the H.E.S.S. data.
However, the H.E.S.S. constraint on $\chi\chi\to q\bar{q}$ becomes
much weaker for $m_{\chi}<300$ GeV.
We point out that this constraint is obtained with NFW profile normalized at  $\rho_{\odot}=0.39~\textrm{GeVcm}^{-3}$. The H.E.S.S. constraint would be slightly less stringent if our adopted normalization $\rho_{\odot}=0.3~\textrm{GeVcm}^{-3}$ is used.
  
Cosmological Constraints on DM annihilation cross section can be obtained from the data of
big-bang nucleosynthesis (BBN) and cosmic microwave background (CMB). In such an analysis, the DM annihilation cross section is assumed to be velocity dependent such that ~\cite{Hisano:2011}
\begin{equation}
\langle \sigma v \rangle=\frac{\langle \sigma v \rangle_0}{\epsilon+(v/v_0)^n},
\end{equation}
where $v_0$ is the DM velocity at the freeze-out temperature, while the values for $\epsilon$ and $n$ depend on specific models. For Sommerfeld enhancement~\cite{Sommerfeld} of the DM annihilation cross section induced by light-scalar exchange, one has $n=1$ and $\epsilon\simeq m_{\phi}/m_{\chi}$ with $m_{\phi}$ the light-scalar mass. The CMB anisotropy can be affected by the energy injection in the recombination epoch due to DM annihilation process such as $\chi\chi\to e^+e^-$ and $\chi\chi\to W^+W^-$. In Fig.~2, we show the upper bound on $\langle \sigma v \rangle$ for $\chi\chi\to e^+e^-$ channel for $n=1$ and $T_{\textrm{KD}}=1\, \textrm{MeV}$
with $T_{\textrm{KD}}$ the kinetic decoupling temperature. This upper bound is inferred from the upper bound on $\langle \sigma v \rangle_0$ such that the resulting CMB power spectrum remains consistent with observations ~\cite{Hisano:2011}. The above upper bound on $\langle \sigma v \rangle_0$ is shown to be sensitive to the parameter $\epsilon$ while the corresponding bound on $\langle \sigma v \rangle$ is insensitive to it.    
It will be interesting to convert the above bound on $\langle \sigma (\chi\chi\to e^+e^-) v \rangle$ into the one on $\langle \sigma (\chi\chi\to \mu^+\mu^-) v \rangle$. However such a conversion is highly model dependent which is beyond the scope of the current work.

Having compared the expected sensitivities of DeepCore detector with other experimental constraints
on various DM annihilation channels, we discuss how the DeepCore constraint on DM annihilation cross section varies with the chosen cone half-angle and threshold energy. We use the channel $\chi\chi\to \mu^+\mu^-$ to illustrate these effects. Fig.~3 shows the required DM annihilation cross
section $\langle\sigma(\chi\chi\to \mu^+ \mu^-)v\rangle$ for a $2\sigma$ detection in five years for different cone half-angle $\psi_{\textrm{max}}$.  
One can see that the constraint on the DM
annihilation cross section gets stronger as $\psi_{\textrm{max}}$
increases from $1^{\circ}$ to $2^{\circ}$. However, the constraint
turns weaker as $\psi_{\textrm{max}}$ increases further. This is due
to the factor $J_{2}(\Delta\Omega)\Delta\Omega$ which depends on the
square of DM density (see Eq.~(\ref{eq4})). The constraint curve
rises with an increasing $\psi_{\textrm{max}}$ for
$\psi_{\textrm{max}} > 2^{\circ}$, since the signal increases slower
than the background does for such a $\psi_{\textrm{max}}$ range. In this figure, we also show the
result for a higher threshold energy $E_\mu^{\textrm{th}}=
50~\textrm{GeV}$ with a cone half-angle
$\psi_{\textrm{max}}=10^{\circ}$ for comparison. This result is
taken from Ref.~\cite{Erkoca:2010} where
$\psi_{\textrm{max}}=10^{\circ}$ is identified as the most optimal
cone half-angle for constraining DM annihilation cross section at
that threshold energy. We note that, for large $m_{\chi}$, lowering
$E_{\mu}^{\textrm{th}}$ from $50$ GeV to $10$ GeV results in more
enhancement on the event rate of atmospheric background than that of
DM annihilation. Hence, the constraint on DM annihilation cross
section is weaker by choosing $E_\mu^{\textrm{th}}=
10~\textrm{GeV}$. On the other hand, for small $m_{\chi}$, lowering
$E_{\mu}^{\textrm{th}}$ enhances more on the event rate of DM
annihilations than that of atmospheric background. For
$m_{\chi}<100$ GeV, one can see that the constraint on DM
annihilation cross section with $E_\mu^{\textrm{th}}=
10~\textrm{GeV}$ is always stronger than that with
$E_\mu^{\textrm{th}}= 50~\textrm{GeV}$. We note that DeepCore constraints on other annihilation channels have similar cone half-angle and threshold energy dependencies. 

Besides studying DeepCore constraints on DM annihilation channels, we also present constraints on DM decay time for $\chi\to b\bar{b}, \ \tau^+ \tau^-$ and $\mu^+\mu^-$ channels. Fig.~4 shows the required DM decay time for a $2\sigma$ detection of neutrino signature in five years for each channel. We have taken $E_\mu^{\textrm{th}}=10~\textrm{GeV}$
and $\psi_{\textrm{max}}=90^{\circ}$.  Non-detection of such a signature would
then exclude the parameter region below the curve at the $2\sigma$
level. For comparison, we also show 
$3\sigma$ limit on $\chi\to \tau^+\tau^-$ from Super-Kamiokande data of upward going muons~\cite{Meade:2009iu}. One can see that the channel $\chi\to \mu^+\mu^-$ requires the smallest decay width to reach the $2\sigma$ detection significance in five years of DeepCore data taking.

\begin{figure}
  \centering
  \includegraphics[width=0.85\textwidth]{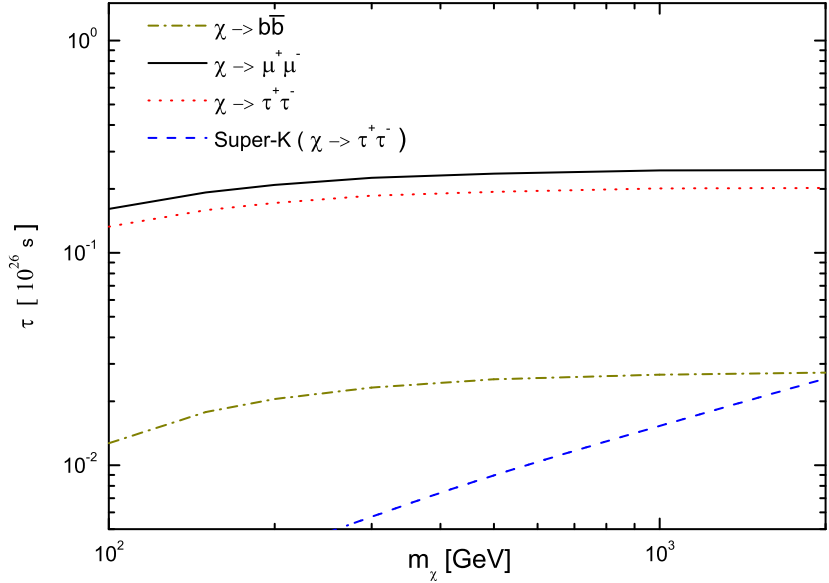}
  \vskip 0.5cm
  \caption{ The dot-dashed line, solid line and doted line are the required DM decay time
  for a $2\sigma$ detection of neutrino signature in five years for $\chi\to b\bar{b}, \ \mu^+ \mu^-$ and $\tau^+\tau^-$ channels, respectively.  The dashed line is the Super-Kamiokande constraint on $\chi\to \tau^+\tau^-$~\cite{Meade:2009iu}.}
  \label{fig:4}
\end{figure}

\begin{figure}
  \centering
  \includegraphics[width=0.85\textwidth]{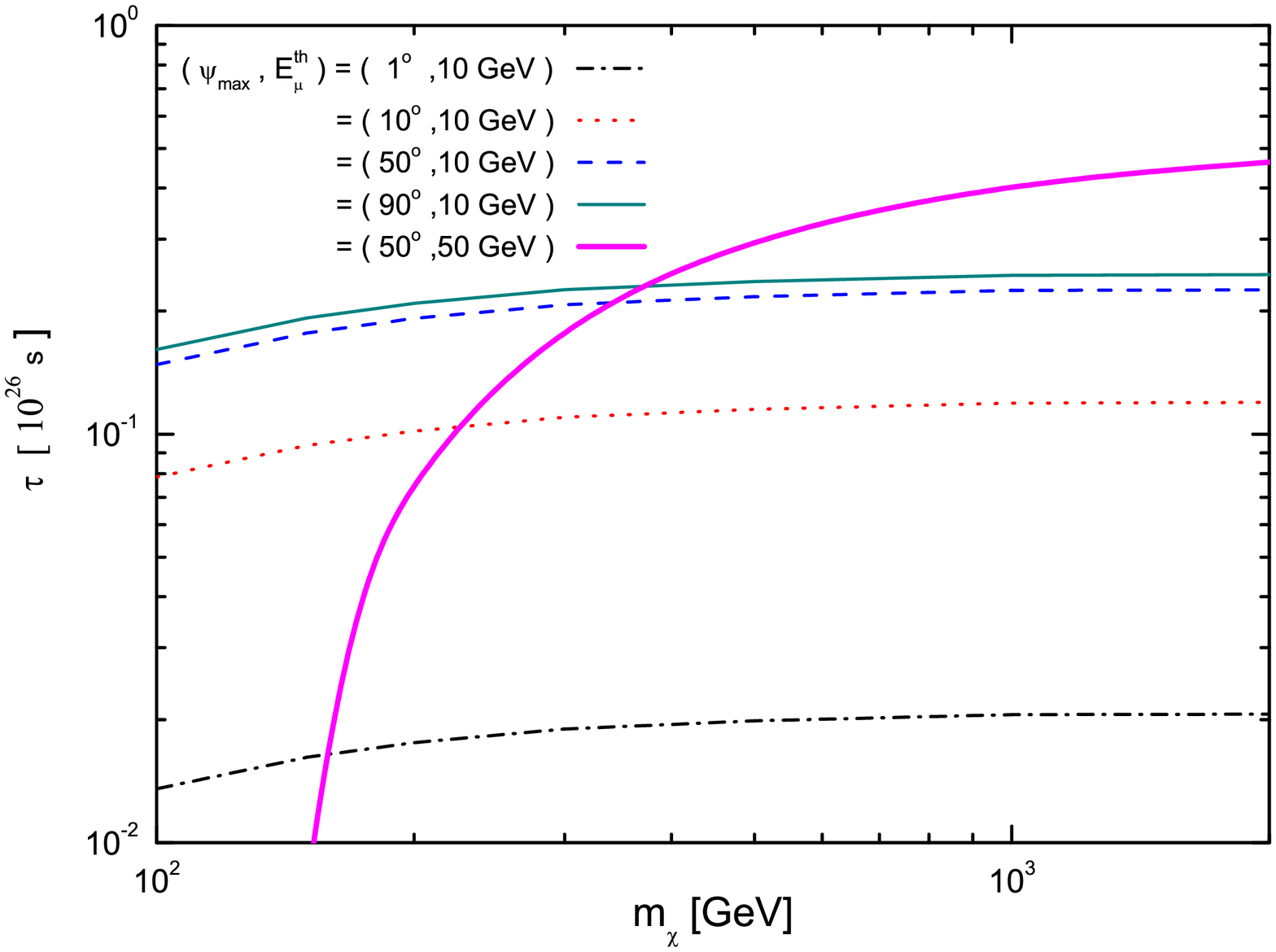}
  \vskip 0.5cm
  \caption{ The required DM decay time $(\chi\rightarrow\mu^{+}\mu^{-})$
  as a function of $m_{\chi}$ such that the neutrino signature from DM decays can be
  detected at the $2\sigma$
  significance in five years.
  Results corresponding to different $\psi_{\rm max}$ are
  presented. For comparison, we also show the result with $E_{\mu}^{\rm th}=50$ GeV
  and $\psi_{\textrm{max}}=50^{\circ}$ \cite{Erkoca:2010}.}
  \label{fig:5}
\end{figure}

Finally we present how the DeepCore constraint on DM decay time varies with the chosen cone half-angle and threshold energy. We use the channel $\chi\to \mu^+\mu^-$ to illustrate these effects. Fig.~5 shows the required DM decay time
$(\chi\rightarrow\mu^{+}\mu^{-})$ as a function of $m_{\chi}$ for
different cone half-angle $\psi_{\textrm{max}}$ such that the
neutrino signature from DM decays can be detected at the $2\sigma$
significance in five years. For DM decays, the curve rises with increasing
$\psi_{\textrm{max}}$ since the event rate of DM signal increases
faster than that of atmospheric background as $\psi_{\textrm{max}}$
increases. For comparison, we show the required DM decay time for a
$2\sigma$ detection in five years with $E_\mu^{\textrm{th}}=
50~\textrm{GeV}$ and $\psi_{\textrm{max}} = 50^{\circ}$. It has been
pointed out in Ref. ~\cite{Erkoca:2010} that $\psi_{\textrm{max}} =
50^{\circ}$ gives the most stringent constraint on DM decay time for
$E_\mu^{\textrm{th}}= 50~\textrm{GeV}$. One can see that the constraint on DM decay time
is strengthen by lowering $E_\mu^{\textrm{th}}$ from $50$ GeV to
$10$ GeV for $m_{\chi}<
300$ GeV.

\section{Summary}

We have calculated the track event rate in IceCube DeepCore array
resulting from muon neutrino flux produced by annihilations and
decays of dark matter in the galactic halo. In this calculation, we
have employed NFW profile for dark matter mass distribution and
consider the channels $\chi\chi\to b\bar{b}, \
\tau^+ \tau^-$, and $\mu^+\mu^-$ for annihilations and the channels
$\chi\to b\bar{b}, \ \tau^+ \tau^-$ and $\mu^+\mu^-$ for decays. We also calculated the track event rate due
to atmospheric background. We compare the signal event rate
with that of the background for $E_{\mu}\geq 10$ GeV.

We have presented sensitivities of IceCube DeepCore array to
neutrino flux arising from dark matter annihilations and decays. For
a given dark matter mass, we evaluated the dark matter annihilation
cross section and dark matter decay time such that a $2\sigma$
detection significance for the above signatures can be achieved by
DeepCore array for a five-year data taking. The DeepCore sensitivities on dark matter annihilation cross section were compared with the constraint obtained from H.E.S.S. gamma ray observations and the constraint derived from the data of CMB power spectrum. Using $\chi\chi\to \mu^+\mu^-$ and $\chi\to \mu^+\mu^-$ as examples, we also presented how DeepCore constraints on dark matter annihilation cross section and dark matter decay time vary with the chosen cone half-angle and threshold energy. We like to point out that our
calculated sensitivities based upon $E_\mu^{\textrm{th}}=
10~\textrm{GeV}$ are significantly more stringent than those obtained by taking $E_\mu^{\textrm{th}}=
50~\textrm{GeV}$ for $m_{\chi}< 100$ GeV in the annihilation channel and
$m_{\chi}<300$ GeV in the decay channel.

\section*{Acknowledgements}
This work is supported by the National Science Council of Taiwan
under Grants No. 099-2811-M-009-055 and 99-2112-M-009-005-MY3, and
Focus Group on Cosmology and Particle Astrophysics, National Center
for Theoretical Sciences, Taiwan.

\end{document}